\documentstyle[prl,aps,tighten]{revtex}
\begin{document}
\draft

\title{Influence of surface irregularities on
barriers for vortex entry in type-II superconductors}

\author{A.Yu.Aladyshkin, A.S.Mel'nikov,
I.A.Shereshevsky and I.D.Tokman}
\address{Institute for Physics of the Microstructures
RAS, 603600, Nizhny Novgorod, GSP-105, Russia}

\date{\today}
\maketitle

\begin{abstract}
A theoretical description of the influence of surface
irregularities (such as wedge-like cracks) on the
Bean-Livingston energy barrier is presented. A careful quantitative
estimate of the field of the first vortex entry $H^*$ into a homogeneous
bulk superconductor with crack-like defects is obtained. This estimate is
based on an exact analytical solution for the Meissner current
distribution in a bulk superconductor with a single two-dimensional
infinitely thin crack.
This thin crack is characterized by the
lowest $H^*$ value, and, thus, appears to be the best gate for vortex
entry.
\end{abstract}

\ \\
\ \\

Experimental  and theoretical
investigations  of influence of
properties of a  sample surface on the Bean-Livingston (BL) barrier
have recently attracted a
great deal of attention (see, e.g., \cite{konczykowski,Burl91,kugel}).
Surface imperfections can result
in a decrease of the critical field $H^*$ of the first vortex
entry into a superconductor:  $H^{*} \sim \gamma H^*_{p}$, where
$H^*_{p}$ is the critical field
corresponding to the BL barrier suppression in the
ideal type-II superconductors, $\gamma<1$.

Provided we neglect the
thermal activation effects (see, e.g., \cite{burlachkov}), the vortex
penetration into a superconductor is possible, if two conditions are
fulfilled:

(i) the current density should be equal to the depairing
(Ginzburg-Landau) current density $j_{GL}$ to form the normal core;

(ii) the energy of the vortex should decrease with an increase of the
distance from the surface, otherwise the vortex could not penetrate into
the bulk.

It is obvious that the properties of the near-surface region
strongly affect these conditions.
In \cite{Burl91,Bass96} the BL barrier suppression has been
analyzed for the case of small ($L<\lambda$) or smooth asperities
($L$ is the depth of defects, $\lambda$ is the
London penetration depth). In this case the suppression of the BL barrier
was shown to be moderate $(\gamma \approx 0.9)$.  We believe the effect
of the barrier suppression to  be the strongest, when the surface defects
have the form of thin deep cracks with dimension $L \stackrel{_>}{_\sim}
\lambda$, oriented across the Meissner current lines, and,
stretched along an external magnetic field $H$.
The existense of such cracks leads to  enhancement of the current density
near the crack edge and, therefore, facilitates creation of vortices and
their further penetration into the bulk \cite{Kopn94,Buzd98}.

The purpose of this paper is to evaluate
the factor $\gamma$ for a single wedge-like crack with $L\gg\lambda$
(the base wedge angle in the plane $(x,y)$ is $\theta_0$, the wedge edge
is parallel to the direction of an external magnetic field $H$,
the $z-$axis is along the wedge edge).  The  equation for the magnetic
field inside a superconductor with a single vortex line reads:

\begin{equation}
\label{eq1}
\Delta B_z-\frac{1}{\lambda^2} B_z=
 -\frac{\phi_0}{\lambda^2}\delta({\bf r}-{\bf
a}),\quad B_z{\Big |}_{ \Gamma}=H.
\end{equation}

\noindent We will use a cylindrical
coordinate system $(r,\theta,z)$ with the origin at the wedge edge,
$\Gamma$ is the contour chosen along  both sides of the wedge ($\theta=0$
and $\theta=2\pi-\theta_0$), the point ($r=a,\theta=\theta_v$) is the
vortex position ($a\ll L$).
For small distances $r\ll \lambda, a\ll
\lambda$ we can neglect the screening effects, and omit second term in
the left-hand side of Eq.~(\ref{eq1}).  Using the conformal mapping
method, we get the magnetic field distribution near the wedge edge:

\begin{equation}
\label{eq2}
B_{z}(r,\theta)=\frac{\Phi_0}{4\pi\lambda^2}
ln\frac{R^{\mu}+R^{-\mu}-2\cos\mu(\theta+\theta_v)}
{R^{\mu}+R^{-\mu}-2\cos\mu(\theta-\theta_v)}
+ H(1-\beta (\frac{r}{\lambda})^{\mu} sin\mu\theta),
\end{equation}

\noindent where
$\mu=\pi/(2\pi-\theta_0)$, $R=r/a$, and $\beta=\beta(\theta_0)$ is the
dimensionless factor.  The Gibbs energy of the vortex per unit length is
given by the expression

\begin{equation}
\label{eq14}
G(a,\theta_v)=\frac{\Phi_0}{4\pi}\left[
- \beta H\Big(\frac{a}{\lambda}\Big)^{\mu} sin(\mu\theta_v)+
\frac{\Phi_0}{4\pi\lambda^2}
ln\Big(\frac{a}{\mu \lambda}\Big) +
\frac{\Phi_0}{4\pi\lambda^2}
ln(2 sin(\mu \theta_v)) + H_{c1} \right].
\end{equation}

\noindent The most
energetically favorable direction of vortex penetration
corresponds to $\theta^*_v=(2\pi-\theta_0)/2$. For this direction the
energy $G(a,\theta^*_v)$ has a maximum at
$a_{max}=\lambda (4\pi\mu\beta\lambda^2 H/\Phi_0)^{-1/\mu}.$
The critical magnetic field $H^*$ corresponding to
vanishing of the energy barrier is

\begin{equation} \label{eq15}
H^*_{\mu}=\frac{1}{\mu \beta}\left (\frac{\xi}{\lambda} \right
)^{1-\mu} H^*_{p}.
\end{equation}

\noindent At fields $H\ge H^*$ the current density near the
wedge edge is of the order of the depairing current density $j_{GL}$,
which is necessary to create a normal core.   Note that maximum
suppression of the BL energy barrier occurs for a thin crack $(\theta_0\ll
1)$, when $\mu=1/2$ and the first vortex entry takes place in an external
magnetic field

\begin{equation}
\label{eq17}
H^*_{0} =\frac{2}{\beta} \frac{1}{\sqrt{\kappa}} H^*_{p},
\end{equation}

\noindent where $\kappa=\lambda/\xi$.

Note that the factor $\beta$ is of fundamental
interest for explanation of the large reduction of the BL barrier
observed experimentally, and can be obtained correctly
only on the basis of a detailed
solution of the screening problem.
The authors of \cite{Buzd98} have
concluded that $\beta$ is of the order of
unity, and so their estimate of the entrance field
${H^*\propto (\xi/\lambda)^{(1-\mu)} H^*_p}$ is valid only within the
arbitrary dimensionless factor. We will
obtain the exact value of $\beta$ for the most interesting case
${\theta_0 \to 0}$, which is necessary to find the ultimate suppression
of the Bean-Livingston barrier.

To obtain the solution it is useful
to note that the London equation (\ref{eq1}) can
be rewritten in the form of the integral equation with auxiliary vortex
sources at the border of a superconductor (see also Ref.\cite{Burl91}):

\begin{equation}
\label{eq4}
\frac{\Phi_0}{2\pi \lambda^2}
\int\limits_0^{\infty} K_0(|r-\rho|/\lambda) n(r) dr = H \quad
(\rho>0),
\end{equation}

\noindent where $K_0(r)$ is the  McDonald function of zero
kind.  Using  the Wiener-Hopf method, we get the following
result:

\begin{equation}
\label{eq7}
n(r)=\frac{2\lambda
H}{\Phi_0} \Big( \frac{e^{-r/\lambda}}{\sqrt{\pi r/\lambda}}+
\frac{2}{\sqrt{\pi}} \int\limits_0^{\sqrt{r/\lambda}} e^{-u^2} du \Big).
\end{equation}

\noindent Using the expression

$$
\frac{\lambda^2}{r} \Big( \left.  \frac{\partial B}{\partial
\theta}\right|_{\theta=0} - \left.  \frac{\partial B}{\partial
\theta}\right|_{\theta=2\pi-\theta_0}\Big)= -\Phi_0 n(r)
\qquad\qquad
$$

\noindent for small $r\ll \lambda$ one can obtain the factor
$\beta=2/\sqrt{\pi}$.

Thus, we obtain the ultimate suppression of the BL barrier. Vortex
penetration starts when an external magnetic field oriented along the
crack achieves the following value:

\begin{equation}
\label{eq18}
H^*_{0} =\frac{\sqrt{\pi}}{\sqrt{\kappa}} H^*_{p}.
\end{equation}

For HTSC $\kappa \sim 10^2$, hence $
H^*_{0}=H^*_{p}\sqrt{\pi/\kappa} \sim 0.18 H_{cm} \sim 3.9 H_{c1}$, where
$H_{c1}$  is the lower critical field.

To summarize, we found out that surface irregularities strongly suppress
the BL barrier for vortex penetration into the bulk and  the rate of
suppression depends on the type of the defects.  The  maximum
suppression occurs for  thin wedge-like cracks oriented parallel to
$H$ and in this case we have found the analytical expression for the
critical field $H^*$ of the first vortex entry.

We are grateful to J.R.Clem, A.A.Andronov, I.L.Maksimov, D.Yu.Vodolazov,
N.B.Kopnin for useful discussions and valuable comments, and  to
A.I.Buzdin for correspondence.  This work was supported, in part, by the
Russian Foundation for  Basic Research (Grant No. 99-02-16188) and by the
International Center for Advanced Studies (INCAS 99-2-03).

\end{document}